\documentclass[aps,twocolumn]{revtex4-2}
\usepackage[utf8]{inputenc}
\usepackage{graphicx}
\usepackage{float}
\usepackage{amsmath}
\usepackage{appendix}
\usepackage{url}

\begin{document}

\title{Ab initio evaluation of the electron-ion energy transfer rate\\
in a non-equilibrium warm dense metal}

\author{Jia Zhang$^{1,2}$, Rui Qin$^{2}$, Wenjun Zhu$^{2}$ and Jan Vorberger$^{1,*}$}

\address {$^{1}$Institute of Radiation Physics, Helmholtz-Zentrum Dresden-Rosendorf, Bautzner Landstraße 400, 01328 Dresden, Germany\\
$^{2}$National Key Laboratory of Shock Wave and Detonation Physics, Institute of Fluid Physics, China Academy of Engineering Physics, Mianyang 621999, China}
\date{\today}

\begin{abstract}
    Electron-ion interactions play a central role for the energy relaxation processes and ultra-fast structure dynamics in laser-heated matter. The accurate prediction of the electron-ion energy exchange in a transient excited two-temperature situation still remains an open and challenging problem even though various theoretical efforts have been made. Here, following our recent work [Zhang $et$ $al$., Materials 15, 1902 (2022)], we take an approach combining finite temperature DFT-MD and corresponding density functional perturbation theory to evaluate the electron-ion coupling factors in the warm dense regime. The use of density-temperature-dependent Eliashberg functions and electron density of states are highlighted. Good agreement of our DFT based results can be observed with recent data by Simoni $et$ $al$. [Phys. Rev. Lett. 122, 205001 (2019)]. For a proof of concept, we use our newly obtained energy transfer rates to discuss temperature equilibration in warm dense aluminum.
\end{abstract}

\maketitle

\section{Introduction}
Warm dense matter (WDM) as an exotic state at the borders of condensed matter and plasma systems has attracted widespread interest over the last decades. As such, WDM is present at extreme conditions of high pressure and temperature. WDM exists in numerous forms in our universe and is produced in the laboratory~\cite{ng2012outstanding,graziani2014frontiers,dorchies2016non,ostrikov2016colloquium,
falk2018experimental,dornheim2018uniform,bonitz2020ab}. 

The complete and deep understanding of WDM is of great importance to high energy density physics~\cite{glenzer2010symmetric}, astrophysics~\cite{guillot1999interiors,heinonen2020diffusion}, planetary science~\cite{nguyen2004melting} as well as material science~\cite{ernstorfer2009formation,waldecker2016electron,cho2016measurement,mo2018heterogeneous,ogitsu2018ab,fernandez2020reduction,grolleau2021femtosecond,chen2021electron,jourdain2021ultrafast,mo2021ultrafast,
lee2021investigation,wu2022ultrafast,nguyen2023direct}. Such extreme states can be created by laser excitation, shock compression, and ion beam irradiation~\cite{falk2018experimental}. 

In the present work, we focus on transient non-equilibrium WDM induced by laser-metal interaction. Before the laser-excited metal reaches a new thermal equilibrium state, a two temperature stage will last up to several tens of pico-seconds. Before this relaxation process, a high electron temperature has already been established on a femto-seconds timescale via electron-electron collisions. The electron-ion coupling leads to the ion subsystem receiving energy from the hot electron subsystem such that electron and ion temperatures equilibrate. This is accompanied by structural changes and non-thermal and thermal effects~\cite{harb2008electronically,ernstorfer2009formation,
giret2014nonthermal,lian2016ab,zhang2016ultrafast,rethfeld2017modelling,zhang2018lattice,
medvedev2020nonthermal,molina2022molecular,arefev2022kinetics,
zier2023pausing}. The connected mechanisms of ultrafast solid-liquid transitions
~\cite{dorchies2016non,mo2018heterogeneous,jourdain2021ultrafast,mo2021ultrafast,
wu2022ultrafast} and non-equilibrium energy transfer~\cite{vorberger2010energy,vorberger2012theory} in this stage are still poorly understood. These processes in the WDM regime share many fundamental aspects with the study of laser excited semi-conductors, laser-matter interaction at ambient conditions~\cite{volkov2019attosecond,schumacher2023ultrafast,de2023ultrafast,
rethfeld2002ultrafast,mueller2013relaxation,bevillon2014free,ono2018thermalization,
silaeva2018ultrafast,lee2021investigation,chen2021electron,grolleau2021femtosecond,riffe2023excitation,giret2011entropy,
sadasivam2017theory,
ono2017nonequilibrium,klett2018relaxation,
sidiropoulos2021probing,
recoules2006effect,yan2016different,
zhang2018lattice,ben2021structural,ono2021lattice,huang2022observation,
tong2021toward,caruso2022ultrafast,han2023prediction,
zier2023pausing}, and non-equilibrium electron-ion interplay\cite{arnaud2013electron,mueller2013relaxation,caro2015adequacy,
brown2016ab,ji2016ab,waldecker2016electron,waldecker2017coherent,
sadasivam2017theory,
waldecker2017momentum,ono2018thermalization,
tamm2018langevin,medvedev2020electron,smirnov2020copper,wingert2020direct,novko2020ultrafast,sidiropoulos2021probing,
zahn2021lattice,tong2021toward,miao2021nonequilibrium,
zahn2022intrinsic,zhang2022energy,
liu2022calibrating,li2022ab,han2023prediction,bomb2023vibrational,riffe2023excitation,
karna2023effect,medvedev2023electron,akhmetov2023electron}. 
Aside from their theoretical description, the diagnostic of such states poses a challenge \cite{glenzer2009x,vorberger2012dynamic,vorberger2018quantum,vorberger2023revealing}. 

There is a big gap between experimental measurement and theoretical modeling on the electron-ion relaxation rate in warm dense metals. Lin $et$ $al$.\cite{lin2008electron} first gave a number of electron-temperature-dependent electron-ion coupling factors for metals with increasing electronic complexity and now the database has become standard~\cite{1}. However, for copper and gold, Cho $et$ $al$.\cite{cho2016measurement} and Mo $et$ $al$.\cite{mo2018heterogeneous} both extracted  much lower values in the corresponding high electron temperature range. 
On the other hand, the latest experimental results by Nguyen $et$ $al$.\cite{nguyen2023direct} have shown stronger electron-ion coupling in warm dense copper than results of Lin $et$ $al$.~\cite{lin2008electron} for relative low electron temperatures. Considering the discrepancies between experiment and simulation, it seems more work is needed on the theoretical front.
 
Nevertheless, much theoretical progress has already been made to calculate the electron-ion coupling strength in hot excited metals. Unfortunately, predictions remain largely inconsistent~\cite{brysk1975thermal,hazak2001temperature,dharma2001results,daligault2008effect,lin2008electron,vorberger2012theory,waldecker2016electron,ogitsu2018ab,fernandez2020reduction,simoni2019first,medvedev2020electron,faussurier2020electron,petrov2021modeling,akhmetov2022effect,medvedev2022electron,nguyen2023direct}. Among these attempts, the two recent works presented by Medvedev $et$ $al$.~\cite{medvedev2020electron} and Simoni $et$ $al$.~\cite{simoni2019first} should be noted. In case of copper, Medvedev $et$ $al$.~\cite{medvedev2020electron} and Simoni $et$ $al$.~\cite{simoni2019first} both predict increasing electron-ion coupling with elevated electron temperature  similar to the data provided by Lin $et$ $al$.~\cite{lin2008electron} but the three curves differ substantially in magnitude.
 As for nickel and iron, the predictions of Lin $et$ $al$.~\cite{lin2008electron} and  Simoni $et$ $al$.~\cite{simoni2019first} actually provide the opposite trend. 
 
 Apart from transition metals, even for a simple metal like aluminum at normal density and melting temperature conditions, various theoretical predictions exhibit large deviations of up to three orders of magnitude~\cite{brysk1975thermal,hazak2001temperature,dharma2001results,daligault2008effect,lin2008electron,vorberger2012theory,waldecker2016electron,simoni2019first}. 

 In order to naturally consider the coexisting effects of electronic excitation and strong ion-ion coupling during the loss of long range order, we adopt a fully ab initio scheme using finite-temperature density functional theory to investigate the electron-ion coupling factor under non-equilibrium WDM conditions~\cite{hohenberg1964inhomogeneous,mermin1965thermal}. In particular, we study the influence of the {\em ion} temperature on the energy transfer rate and the relaxation. 

 In this work, we will first show a brief derivation for the electron-ion coupling factor and provide a brief introduction of the two temperature model and give the computational procedures and details in our simulations. In the third part, we display our data and compare our electron-ion energy transfer rates with Simoni $et$ $al$.~\cite{simoni2019first} as well as show the heat capacities for electrons and ions under extreme conditions to predict the temperature relaxation. 
 Lastly, we conclude our results and give an outlook on the related topics. All the original data are available in the appendix.

\section{Methodology}
\subsection{Electron-ion coupling factor}
We adopt a microscopic formula derived under the framework of the Bloch-Boltzmann-Peierls equation as a starting point~\cite{allen1987theory}, the detailed derivation can be found elsewhere~\cite{wang1994time,zhang2022energy}. The general expression for the electron-ion energy exchange rate applied to metal systems is as follows
\begin{equation}
\begin{split}
 Z_{ei}(\rho,T_{e},T_{i})&=2\pi
 N_{c}g[\mu(\rho,T_{e},T_{i})]\\
 &\times\int_{0}^{\infty}\!\int_{-\infty}^{\infty}\!\int_{-\infty}^{\infty}\!
 d\omega d\varepsilon d\varepsilon'
 \;(\hbar\omega)\\
&\times\alpha^{2}F(\varepsilon,\varepsilon',\omega,\rho,T_{e},T_{i})\\
 &\times[f(\varepsilon,T_{e})-f(\varepsilon',T_{e})]\\
 &\times[n_{B}^{i}(\omega,T_{i})-n_{B}^{e}(\omega,T_{e})]\\
 &\times\delta(\varepsilon-\varepsilon'+\hbar\omega),
\end{split}
\label{zep1}
\end{equation}
where $N_{c}$ is volume of the simulation cell, $g[\mu(\rho,T_{e},T_{i})]$ is the average electron density of states at the chemical potential, ${f(\varepsilon,T_{e})}$ and ${n_{B}(\omega,T_{i})}$ are the Fermi and Bose distribution function featuring the electron temperature $T_e$ and the ion temperature $T_i$, respectively. The delta function $\delta(\varepsilon-\varepsilon'+\hbar\omega)$  is used to fulfill the energy conservation between the initial and the final state of the scattering electron. $\alpha^{2}F(\varepsilon,\varepsilon^{'},\omega,\rho,T_{e},T_{i})$ is the electron-ion Eliashberg function 
\begin{equation}
\begin{split}
\alpha^{2}F(\varepsilon,\varepsilon',\omega,\rho,T_{e},T_{i})&=\frac{2}{\hbar N_{c}^{2}g[\mu(\rho,T_{e},T_{i}]}\\
    &\times \sum_{k k'}|M_{k k'}^{q}( \rho,T_{e},T_{i})|^{2}\\
    &\times\delta(\omega-\omega_{q})
    \delta(\varepsilon-\varepsilon_{k})\delta(\varepsilon'-\varepsilon_{k'}),
\end{split}
\label{zep2}
\end{equation}  
in which, $M_{k k'}^{q}( \rho,T_{e},T_{i})$ stands for the electron-ion transition matrix element.

Based on the different energy scales of electrons (eV) and phonons (meV), the formula (\ref{zep1}) can be simplified by introducing two approximations. The first concerns the Eliashberg function $\alpha^{2}F(\varepsilon,\varepsilon',\omega,\rho,T_{e},T_{i})$, following Wang $et$ $al$. we obtain~\cite{wang1994time}
\begin{equation}
\begin{split}  \alpha^{2}F(\varepsilon,\varepsilon',\omega,\rho,T_{e},T_{i})
   &=\frac{g^2(\varepsilon,\rho,T_{e},T_{i})}{g^2[\mu(\rho,T_{e},T_{i})]}\alpha^{2}F(\omega,\rho,T_{e},T_{i}),
\end{split}
\label{zep3}
\end{equation} 
in which the assumption is made that only electronic states around the chemical potential are contributing to the energy exchange. And the second approximation simplifies the factor with the difference between Fermi distribution functions 
\begin{equation}
\begin{split}  
    f(\varepsilon,T_{e})-f(\varepsilon',T_{e})=-\hbar\omega\frac{\partial f(\varepsilon,T_{e})}{\partial\varepsilon},
\end{split}
\label{zep4}
\end{equation} 
By inserting expressions (\ref{zep3}) and (\ref{zep4}) into the general formula (\ref{zep1}), we can get the final form for electron-ion coupling factor by dividing it by the temperature difference:

\begin{equation}
\begin{split} 
        G_{ei}(\rho,T_{e},T_{i})&=
        \frac{2\pi N_{c}}{g[\mu(\rho,T_{e},T_{i})](T_{e}-T_{i})}\\
        &\times\int_{-\infty}^{\infty}g^{2}(\varepsilon,\rho,T_{e},T_{i})\frac{\partial f(\varepsilon,T_{e})}{\partial\varepsilon}d\varepsilon\\
        &\times\int_{0}^{\infty}(\hbar\omega)^{2}\alpha^{2}F(\omega,\rho,T_{e},T_{i})d\omega\\
        &\times[n_{B}^{e}(\omega,T_{e})-n_{B}^{i}(\omega,T_{i})].
\end{split}
\label{zep5}
\end{equation}

\subsection{Two-temperature model}
To study the non-equilibrium energy exchange and temperature evolution in the laser-created WDM regime, 
the simple two-temperature model is still a powerful tool to predict the energy transfer between electrons and ions due to the fact that the essence of the main micro-physical processes during the sophisticated relaxation processes are described~\cite{anisimov1967effect,anisimov1974electron,allen1987theory}.
 In our work, we adopt two coupled equations without laser source term as follows
\begin{equation}
\begin{split} 
          C_{e}\frac{\partial T_{e}}{\partial t}&=G_{ei}(T_{i}-T_{e})\\
          C_{i}\frac{\partial T_{i}}{\partial t}&=G_{ei}(T_{e}-T_{i}).
\end{split}
\label{zep6}
\end{equation} 
in which, all the heat capacities $C_{e}(\rho,T_{e},T_{i})$, $C_{i}(\rho,T_{e},T_{i})$ as well as the energy transfer rate $G_{ei}(\rho,T_{e},T_{i})$ are fully temperature dependent and determined via DFT. The heat capacities are calculated by symmetric finite difference quotient from the internal energies.

\subsection{First-principles calculation scheme}

In order to generate the possible atomic configurations encountered in the laser-irradiated aluminum under WDM conditions, we perform ab initio molecular dynamics simulations to monitor the evolution of ions using the package VASP~\cite{kresse1996efficient,kresse1999ultrasoft}. 32 aluminum atoms in a periodic cubic box at three different densities ($\rho=6.4$~g/$cm^{3}$, $\rho=2.7$~g/$cm^{3}$, $\rho=2.35$~g/$cm^{3}$) are prepared to be simulated under the two different target temperatures. A time step of 0.2~fs was used in the MD simulations. In our implementation, the electron temperature was determined by Fermi smearing varying from 0.5~eV to 2.0~eV and the ion temperature was controlled by a Nose thermostat to be at 0.1~eV or 0.3~eV, resp.. The exchange-correlation functional is taken in the generalized gradient approximation (GGA)~\cite{perdew1996generalized}. We use a projector augmented wave (PAW) pseudo-potential, in which the valence state is $3s^{2}3p^{1}$. The plane wave energy cut-off is set to 353~eV for all the calculations. In the DFT-MD simulations, we take 6000-8000 timesteps to relax the ions and then a further 4000 timesteps at the relaxed conditions from which the ionic snapshots are chosen. 

The average electronic density of states under different electron and lattice temperatures are obtained by solving the finite temperature Kohn-Sham equation~\cite{mermin1965thermal,kohn1965self}. Ten configurations  randomly selected from the corresponding DFT-MD runs are used as initial input. Similarly, the average temperature-dependent Eliashberg functions and related phonon density of states are calculated using the framework of the linear response formalism~\cite{gonze1997dynamical,gonze1997first,savrasov1996electron,baroni2001phonons}. All these simulations were performed using the open-source software ABINIT~\cite{gonze2009abinit,gonze2016recent}. To model the energy transfer rates between electrons and ions, we adopt a GGA norm-conserving pseudo-potential with the same number of valence electrons as used in the DFT-MD simulations~\cite{perdew1996generalized}. A plane wave energy cutoff of 25~Ha (680~eV) is used here. During the calculations of the electronic structure, the tetrahedron method is used featuring a k-point grid of up to $24\times24\times24$. 
In the high density case, we set the number of bands to a maximum of 240 at elevated electron temperatures. For the normal and low density cases, we increased the number of bands to 300 and to 350, respectively. 

 To evaluate the temperature-dependent electron–ion transition matrix elements, a combination of an unshifted k-point grid featuring $4\times4\times4$ with the q-point grid of $2\times2\times2$ is used. Due to our supercell containing 32 ions, 96 separate perturbation computations are required to construct the final Eliashberg functions.



\section{Results and discussions}
\begin{figure*}[!htbp]
\centering
\includegraphics[width=0.96\textwidth]{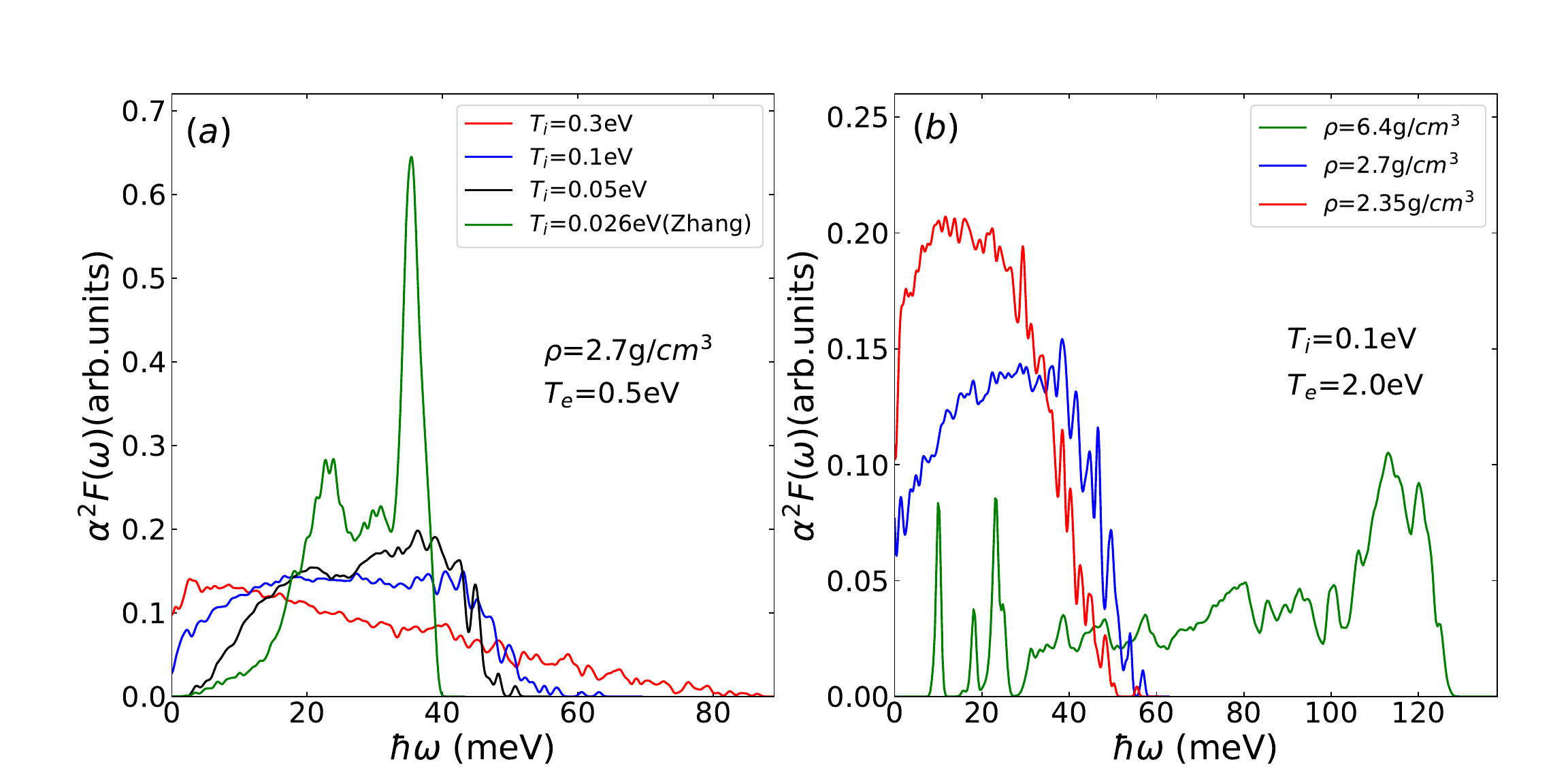}
\caption{The behaviour of the electron-ion spectrum of aluminum under WDM conditions. (a)The average Eliashberg functions for three elevated ion temperatures at normal density under low electronic excitation. The green curve is for an ideal lattice taken from our recent work~\cite{zhang2022energy}. (b)The average Eliashberg functions for three different densities at intermediate ion temperature under high electronic excitation. Further density-temperature-dependent Eliashberg functions are shown in panels a) of Figs.~\ref{Figure_4.pdf}, \ref{Figure_5.pdf}, \ref{Figure_6.pdf}, \ref{Figure_7.pdf}, \ref{Figure_8.pdf} \& \ref{Figure_9.pdf} in Appendix~\ref{A}.}
\label{Figure_1.pdf}
\end{figure*}
In order to study the microscopic information of the electron-ion interaction, we extract the average Eliashberg function for aluminum under transient non-equilibrium conditions. The results for the Eliashberg functions at various conditions are presented in Fig.~\ref{Figure_1.pdf}. We can see that the ion temperature has a great impact on the shape of the Eliashberg function. Specifically, with increasing ion temperature, the transverse and longitudinal peaks of the spectral function slowly disappear and the entire Eliashberg function is broadened. This is due to anharmonic effects in the lattice at only slightly elevated ion temperatures and due to melting at higher ion temperatures. We can observe a huge redistribution of weight from longitudinal to transverse mode as well. 

The variation of the Eliashberg function with density (at high electron temperature) is shown in panel b) of Fig.~\ref{Figure_1.pdf}. We find a broadening  and thus a shift of spectral weights to higher energies at elevated density. As the increase in density leads to solidification of the system, distinct excitation peaks can be observed for the highest density. On the contrary, the feature of the Eliashberg functions near zero frequency for the $\rho=2.35$~g/$cm^{3}$ (red) and $\rho=2.7$~g/$cm^{3}$ (blue) curves show the influence of the short range ordered liquid state of warm dense aluminum on the electron-ion correlations. To rigorously check the effect of electronic excitations, we applied different electron temperatures in the calculation of Eliashberg function of all cases, the total results can be seen from panels a) of Figs.~\ref{Figure_4.pdf}, \ref{Figure_5.pdf}, \ref{Figure_6.pdf}, \ref{Figure_7.pdf}, \ref{Figure_8.pdf} \& \ref{Figure_9.pdf} in Appendix~\ref{A}. It seems that the influence of the electron temperature over the considered electron temperature range is not large but also should not be neglected~\cite{zhang2022energy}.

\begin{figure*}[!htbp]
\centering
\includegraphics[width=0.96\textwidth]{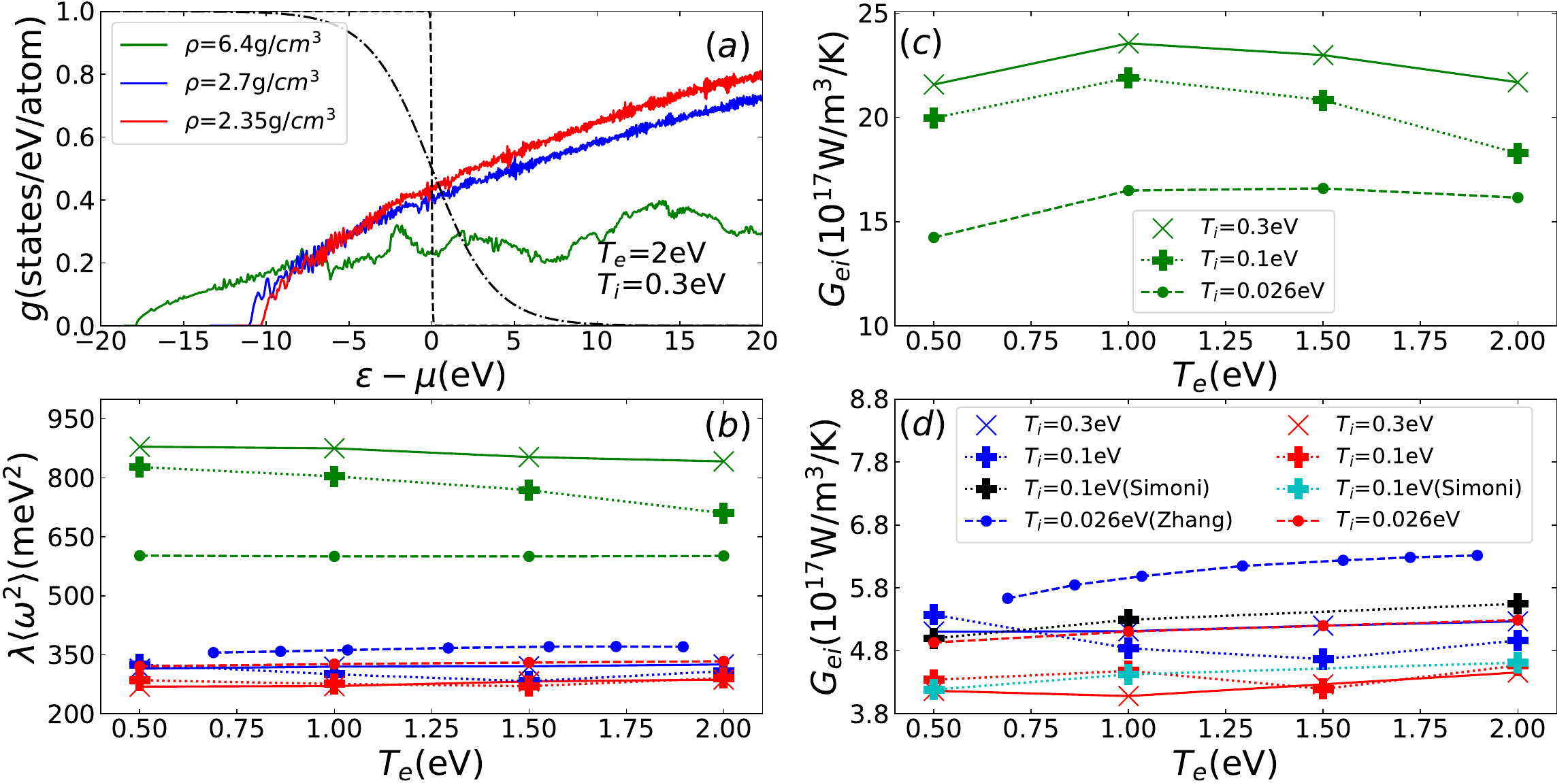}
\caption{The electronic structure and energy transfer rate of aluminum under WDM conditions. (a)The electron density of states for three different densities at high temperatures. The two black dashed lines are the Fermi distribution functions at electron temperatures 0.026~eV and 2.0~eV. (b)The second moment of the Eliashberg function with increasing electron temperature as well as elevated ion temperatures for the corresponding three densities (color codes and symbols as in panels c) \& d)). (c)\&(d)The density-temperature-dependent electron-ion coupling factors. We particularly compare our present results with Simoni $et$ $al$.~\cite{simoni2019first}. The blue dashed lines in panels b) \& d) are adopted from our recent work~\cite{zhang2022energy}. All the detailed information for the associated physical quantities can be accessed in Appendix~\ref{B}.}
\label{Figure_2.pdf}
\end{figure*}

To obtain the final electron-ion energy exchange rate [see Eq. (\ref{zep5})], the average electron density of state at the chemical potential (see panel c) of Figs.~\ref{Figure_10.pdf}, \ref{Figure_11.pdf} \& \ref{Figure_12.pdf} in  Appendix~\ref{B}) is also needed in addition to the Eliashberg function. It can be extracted from the electron density of states combined with the Fermi distribution. From Fig.~\ref{Figure_2.pdf} a), we find that the electron density of states broadens with increasing density. In addition, the typical, distinct peaks present in the DOS of the solid are absent in the fluid DOS, which is even more free-electron like.

The second moment of the Eliashberg function, see Fig.~\ref{Figure_2.pdf} b), is also more influenced by the density and ion temperature than by the electron temperature. This is consistent with the behaviour of the Eliashberg functions at different conditions as shown in  Appendix~\ref{A}, see panels a) of Figs.~\ref{Figure_4.pdf}, \ref{Figure_5.pdf}, \ref{Figure_6.pdf}, \ref{Figure_7.pdf}, \ref{Figure_8.pdf} \& \ref{Figure_9.pdf}. The same trend is seen in the electron-ion coupling factor in Fig.~\ref{Figure_2.pdf}, panels c) \& d). For the highest density, the values of the electron temperature-dependent electron-ion coupling factors are increasing with elevated ion temperature; but for the low density cases, the opposite variation is observed. For the WDM conditions at $\rho=2.7$~g/$cm^{3}$ and $\rho=2.35$~g/$cm^{3}$ (see also in Appendix~\ref{B} the panels f) of Figs.~\ref{Figure_11.pdf} \& \ref{Figure_12.pdf}), our results are very close to data by Simoni $et$ $al$.~\cite{simoni2019first}. Small deviations at $T_{e}=1.5$~eV stem from our approximation of the expansion of the Eliashberg function around the chemical potential.

Simoni $et$ $al$.~\cite{simoni2019first} established a microscopic expression for the electron-ion energy transfer rate in the framework of Langevin-like dynamics.
Then so called quantum friction described by the electron-ion force operator can be used to 
calculate the electron-ion coupling. In our approach as well as in Simoni's, it is necessary to conduct ab initio molecular dynamics simulation to generate typical configurations and then extract the microscopic information on the electron-ion interaction. In this context, the Eliashberg function and quantum friction both provide complementary insights into the non-equilibrium electron-ion energy exchange. The good agreement between our results and Simoni $et$ $al$.~\cite{simoni2019first} indicates that our formula based on the physical picture of an electron-phonon interaction is still a reasonable approximation when applied to WDM situations at moderate ion temperature featuring strong short range ordering of the ions.

\begin{figure*}[!htbp]
\centering
\includegraphics[width=0.96\textwidth]{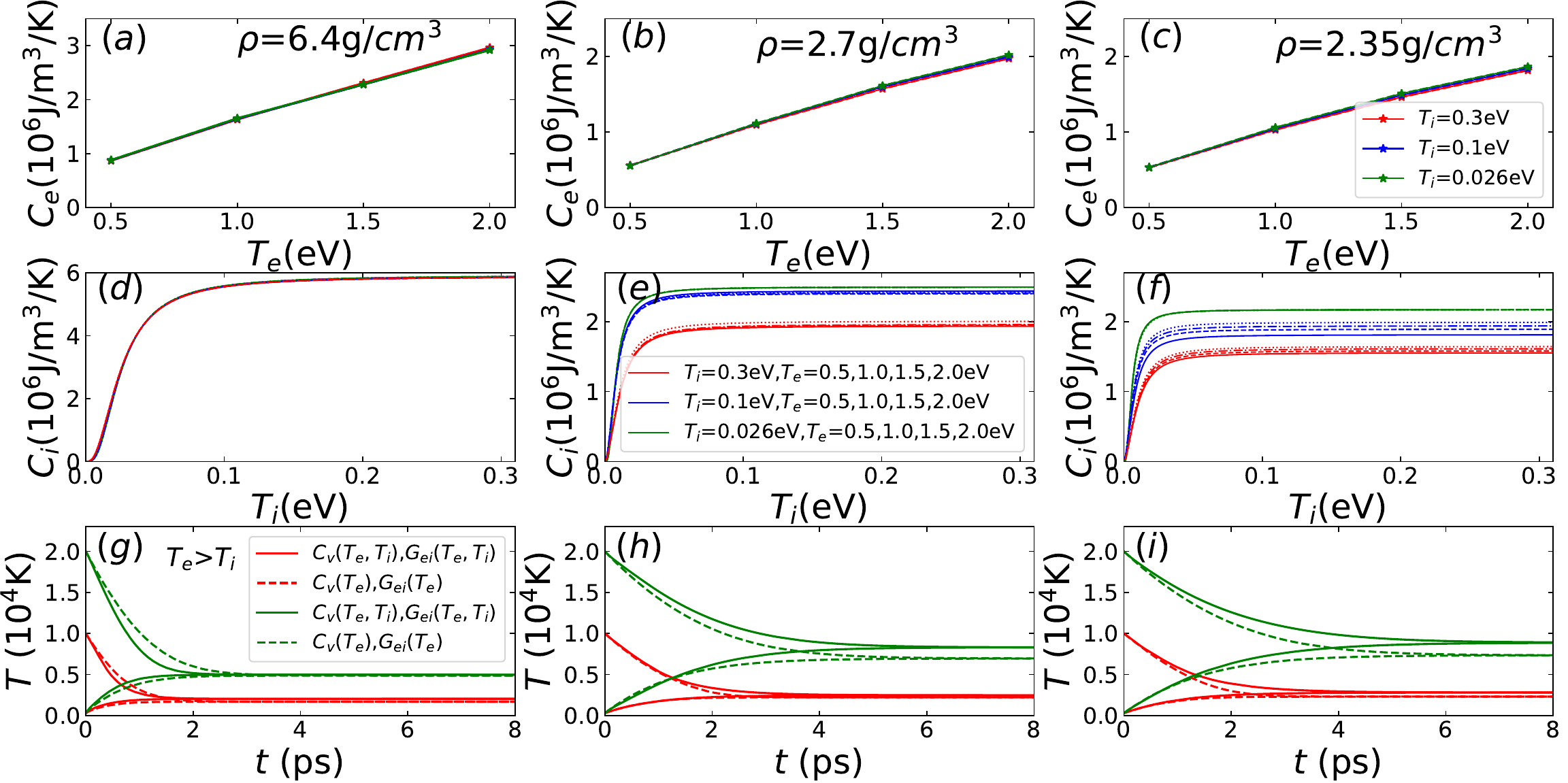}
\caption{The thermal parameters in the two-temperature model and energy relaxation of aluminum under WDM conditions. (a), (b) \& (c) The electron heat capacities as a function of electron temperature at elevated ion temperatures for three different densities, resp.. (d), (e) \& (f) The lattice heat capacities as a function of ion temperature at elevated electron temperatures for three different densities, resp.. The line styles denote different electron temperature. (g), (h) \& (i) The temperature equilibration between electron and ion subsystems. We solve the two-temperature model using fully temperature-dependent coefficients. In comparison, the case for only electron-temperature-dependent parameters is also shown.}
\label{Figure_3.pdf}
\end{figure*}
In the two-temperature model, the heat capacities for electrons and ions are the key parameters to control the profile of the temperature evolution. Based on the electron DOS (shown in Appendix~\ref{B}, Figs.~\ref{Figure_10.pdf}, \ref{Figure_11.pdf} \& \ref{Figure_12.pdf}) and phonon DOS (shown in Appendix~\ref{A}, Figs.~\ref{Figure_4.pdf}, \ref{Figure_5.pdf},  \ref{Figure_6.pdf}, \ref{Figure_7.pdf}, \ref{Figure_8.pdf} \& \ref{Figure_9.pdf}), the corresponding heat capacities are calculated. We find that the electron heat capacity is not sensitive to the effect of ion temperature from Fig.~\ref{Figure_3.pdf} a), b) \& c). On the contrary, in Fig.~\ref{Figure_3.pdf} d), e) \& f), both the ion and electron temperatures have an obvious impact on the lattice heat capacities at normal and relative low density. 

We apply this fully temperature-dependent method to solve the two temperature model. As comparison, an approach with only electron-temperature-dependency is also used. From Fig.~\ref{Figure_3.pdf} g), we find that the relaxation time is shorter in the highest density situation when the fully temperature dependent model is used. For the low density cases, we observe that the equilibration time is longer and the final equilibrium temperature is higher, see Fig.~\ref{Figure_3.pdf} h) \& i).
Given the profound effect of temperature and density on heat capacity and energy transfer rate, it indicates that the temperature relaxation in the WDM regime is a complicated, not yet completely understood, process. 

\section{Summary and outlook}
We have investigated the rate of energy exchange and temperature relaxation between electrons and ions for the simple metal aluminum under WDM conditions and excited by ultrafast laser irradiation. By using finite temperature DFT-MD and the associated linear response method, we paid special attention into calculating the average electron density of states and  Eliashberg functions for all considered temperatures as well as densities to accurately estimate the density and temperature dependent electron–ion coupling strengths $G_{ei}$($\rho$,$T_{e}$,$T_{i}$). 

At ambient and relative low densities for aluminum, our results for the electron-ion coupling factors show a good match with results by Simoni $et$  $al$.~\cite{simoni2019first}. It implies that the Eliashberg function method is still valid under such extreme conditions.

For high lattice temperature (or generally high ion temperatures), our prediction may be invalid due to  anharmonic effects or unstable phonons. To overcome the breakdown of the electron-phonon picture in the energy transfer, a different quantum many-body theoretical framework for the energy exchange rate in WDM conditions is needed. Recently, we have proposed a promising scheme using the non-equilibrium green's function technique and we hope our approach will provide a new perspective of electron-ion energy exchange in this transient exotic state in the future~\cite{vorberger2010energy,vorberger2012theory}. 

Apart from the TTM, the energy relaxation and melting dynamics in laser-heated metals can be modeled by two-temperature molecular dynamics~\cite{ivanov2003combined,duffy2006including,
zhang2018lattice}. The extracted static structure factor at WDM conditions in 2T-MD simulation can be directly compared with ultrafast diffraction experiment\cite{mo2018heterogeneous,molina2022molecular,arefev2022kinetics}. Using the key parameters as determined here by ab initio methods, it will provide a possibility to test and verify our theoretical model and eventually improve our understanding of non-equilibrium WDM.

\begin{acknowledgments}
This work was supported by China Scholarship Council.  We gratefully acknowledge the CPU time granted on the  hemera cluster of the High-Performance Computing Center at HZDR and on a Bull Cluster at the Center for Information Services and High-Performance Computing (ZIH) at Technische Universität Dresden.
\end{acknowledgments}

\appendix
\section{The average Eliashberg functions and corresponding phonon density of states }
\label{A}
\begin{figure}[H]
\centering
\includegraphics[width=0.48\textwidth]{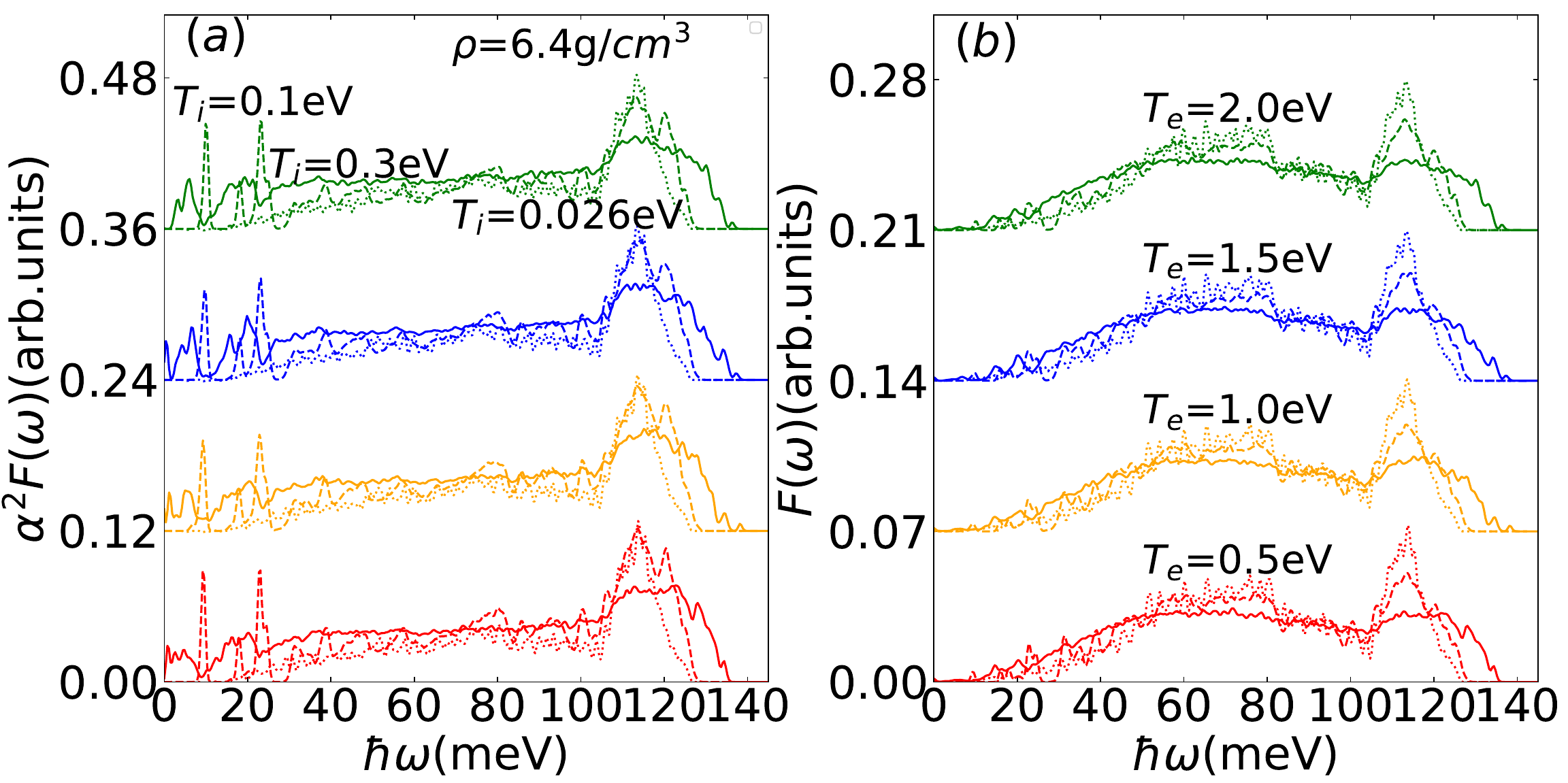}
\caption{(a)Electron and ion temperature-dependent Eliashberg functions and (b)associated phonon density of states of aluminum at $\rho=6.4$~g/$cm^{3}$. The green, blue, orange and red curves stand for the electron temperatures of 2.0~eV, 1.5~eV, 1.0~eV and 0.5~eV, resp.. The solid, dashed and dotted lines correspond to the ion temperatures of 0.3~eV, 0.1~eV and 0.026~eV, rep.. }
\label{Figure_4.pdf}
\end{figure}

\begin{figure}[H]
\centering
\includegraphics[width=0.48\textwidth]{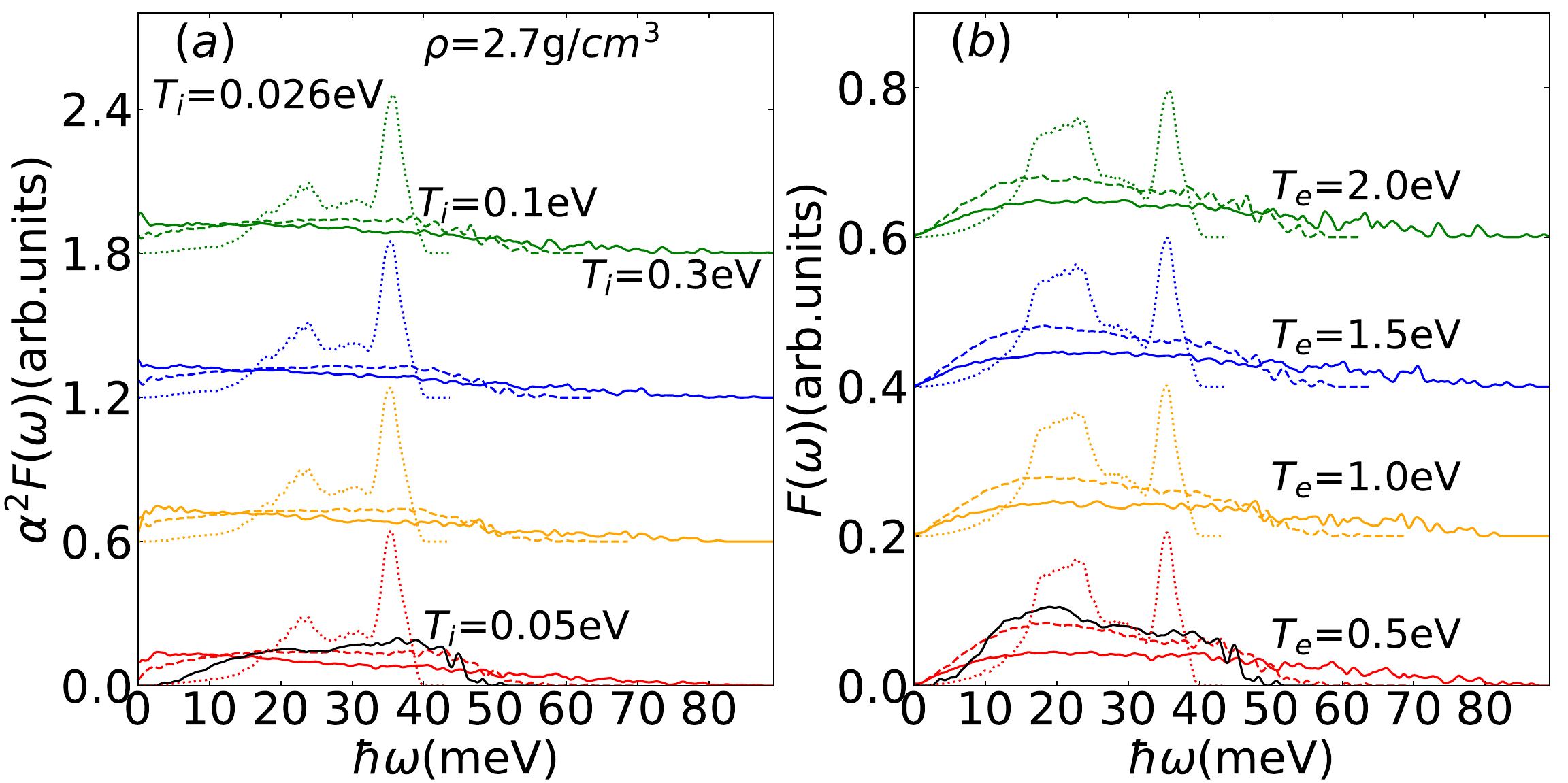}
\caption{(a)Electron and ion temperature-dependent Eliashberg functions and (b)associated phonon density of states of aluminum  at $\rho=2.7$~g/$cm^{3}$. The green, blue, orange and red curves stand for the electron temperature 2.0~eV, 1.5~eV, 1.0~eV and 0.5~eV, resp.. The solid, dashed and doted lines correspond to the ion temperature 0.3~eV, 0.1~eV and 0.026~eV, rep.. All  doted lines are taken from our recent work~\cite{zhang2022energy}. The two black curves are shown here under the condition of $T_{i}=0.05$~eV and $T_{e}=0.5$~eV.}
\label{Figure_5.pdf}
\end{figure}

\begin{figure}[H]
\centering
\includegraphics[width=0.48\textwidth]{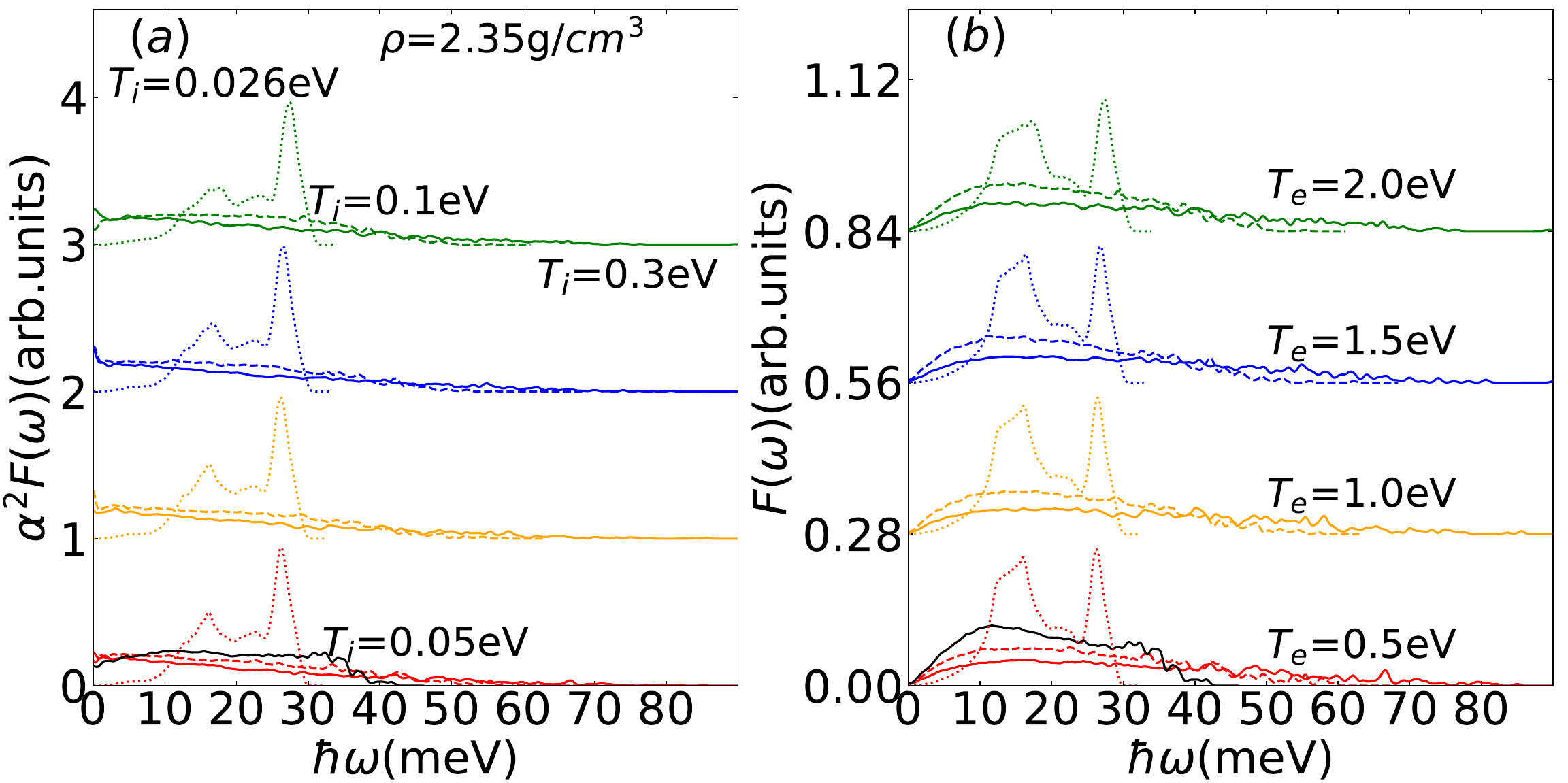}
\caption{(a)Electron and ion temperature-dependent Eliashberg functions and (b)associated phonon density of states of aluminum  at $\rho=2.35$~g/$cm^{3}$. The green, blue, orange and red curves stand for the electron temperature 2.0~eV, 1.5~eV, 1.0~eV and 0.5~eV, resp.. The solid, dashed and doted lines correspond to the ion temperature 0.3~eV, 0.1~eV and 0.026~eV, rep.. The two black curves are shown here under the condition of $T_{i}=0.05$~eV and $T_{e}=0.5$~eV.}
\label{Figure_6.pdf}
\end{figure}

\begin{figure}[H]
\centering
\includegraphics[width=0.48\textwidth]{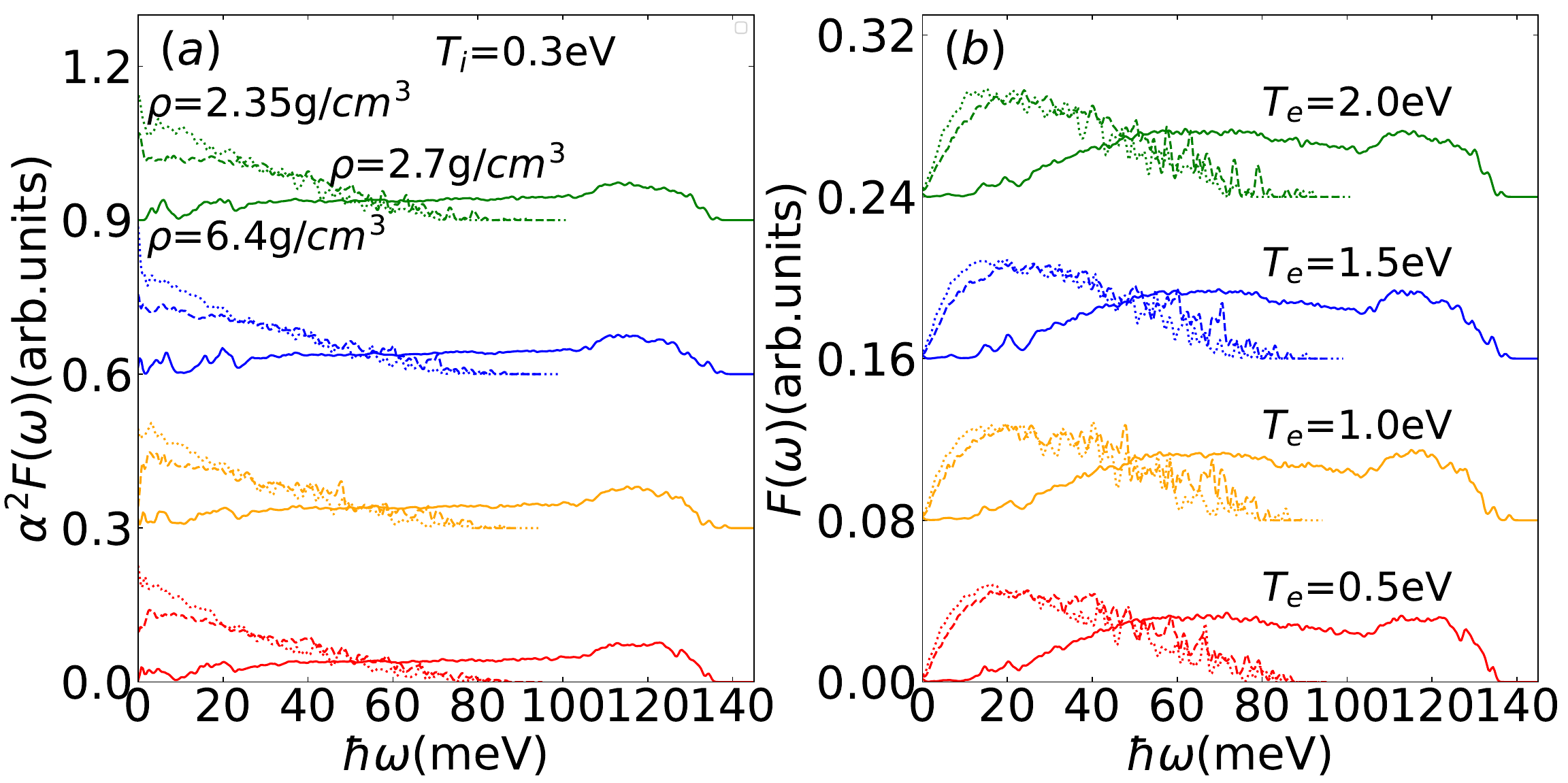}
\caption{(a)Density and electron temperature-dependent Eliashberg functions and (b)associated phonon density of states of aluminum  at $T_{i}=0.3$~eV. The green, blue, orange and red curves stand for the electron temperatures of 2.0~eV, 1.5~eV, 1.0~eV and 0.5~eV, resp.. The solid, dashed and doted lines correspond to the densities $\rho=6.4$~g/$cm^{3}$, $\rho=2.7$~g/$cm^{3}$ and $\rho=2.35$~g/$cm^{3}$, rep.. }
\label{Figure_7.pdf}
\end{figure}
\begin{figure}[H]
\centering
\includegraphics[width=0.48\textwidth]{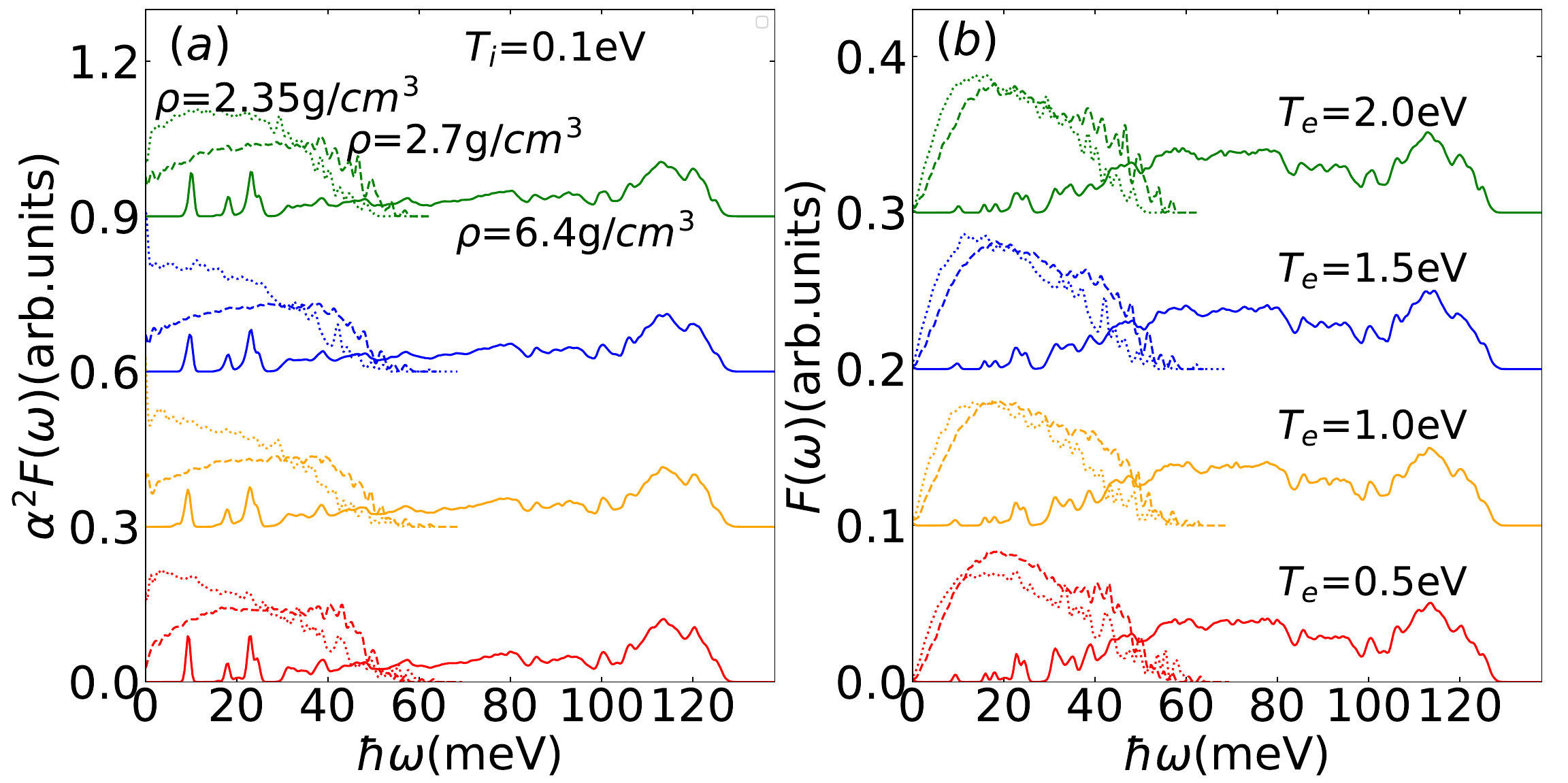}
\caption{(a)Density and electron temperature-dependent Eliashberg functions and (b)associated phonon density of states of aluminum  at $T_{i}=0.1$~eV. The green, blue, orange and red curves stand for the electron temperature 2.0~eV, 1.5~eV, 1.0~eV and 0.5~eV, resp.. The solid, dashed and doted lines correspond to the density $\rho=6.4$~g/$cm^{3}$, $\rho=2.7$~g/$cm^{3}$ and $\rho=2.35$~g/$cm^{3}$, rep.. }
\label{Figure_8.pdf}
\end{figure}

\begin{figure}[H]
\centering
\includegraphics[width=0.48\textwidth]{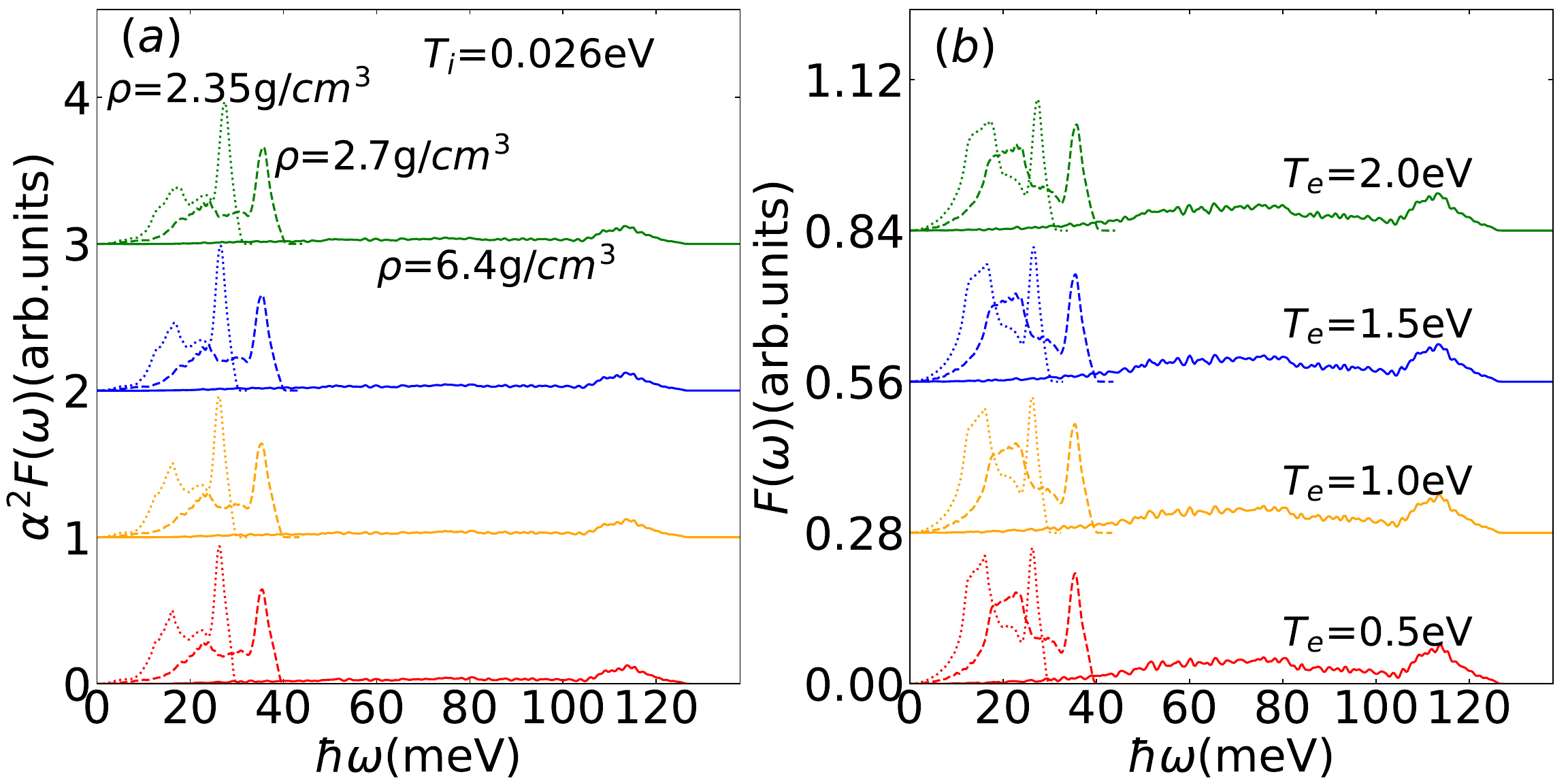}
\caption{(a)Density and electron temperature-dependent Eliashberg functions and (b)associated phonon density of states of aluminum  at $T_{i}=0.026$~eV. The green, blue, orange and red curves stand for the electron temperature 2.0~eV, 1.5~eV, 1.0~eV and 0.5~eV, resp.. The solid, dashed and doted lines correspond to the density $\rho=6.4$~g/$cm^{3}$, $\rho=2.7$~g/$cm^{3}$ and $\rho=2.35$~g/$cm^{3}$, rep.. All dashed curves are adopted from our recent work~\cite{zhang2022energy}.}
\label{Figure_9.pdf}
\end{figure}

\section{Electron-ion energy transfer rates}
\label{B}
The chemical potential can be calculated by equation $N_{e}=\int_{\infty}^{\infty}g(\varepsilon,\rho,T_{e},T_{i}) f(\varepsilon,\mu(\rho,T_{e},T_{i}),T_{e})d\varepsilon$, in which $N_{e}$ equals to the total number of valence electrons. The second moment of the Eliashberg function and the electron-ion coupling constant can be obtained by formulas
$\lambda\langle w^{2}\rangle(\rho,T_{e},T_{i})=2\int_{0}^{\infty}\omega\alpha^{2}F(\omega,\rho,T_{e},T_{i})d\omega$ and
$\lambda(\rho,T_{e},T_{i})=2\int_{0}^{\infty}\frac{\alpha^{2}F(\omega,\rho,T_{e},T_{i})}{\omega}d\omega$, respectively.

\begin{figure}[!htbp]
\centering
\includegraphics[width=0.48\textwidth]{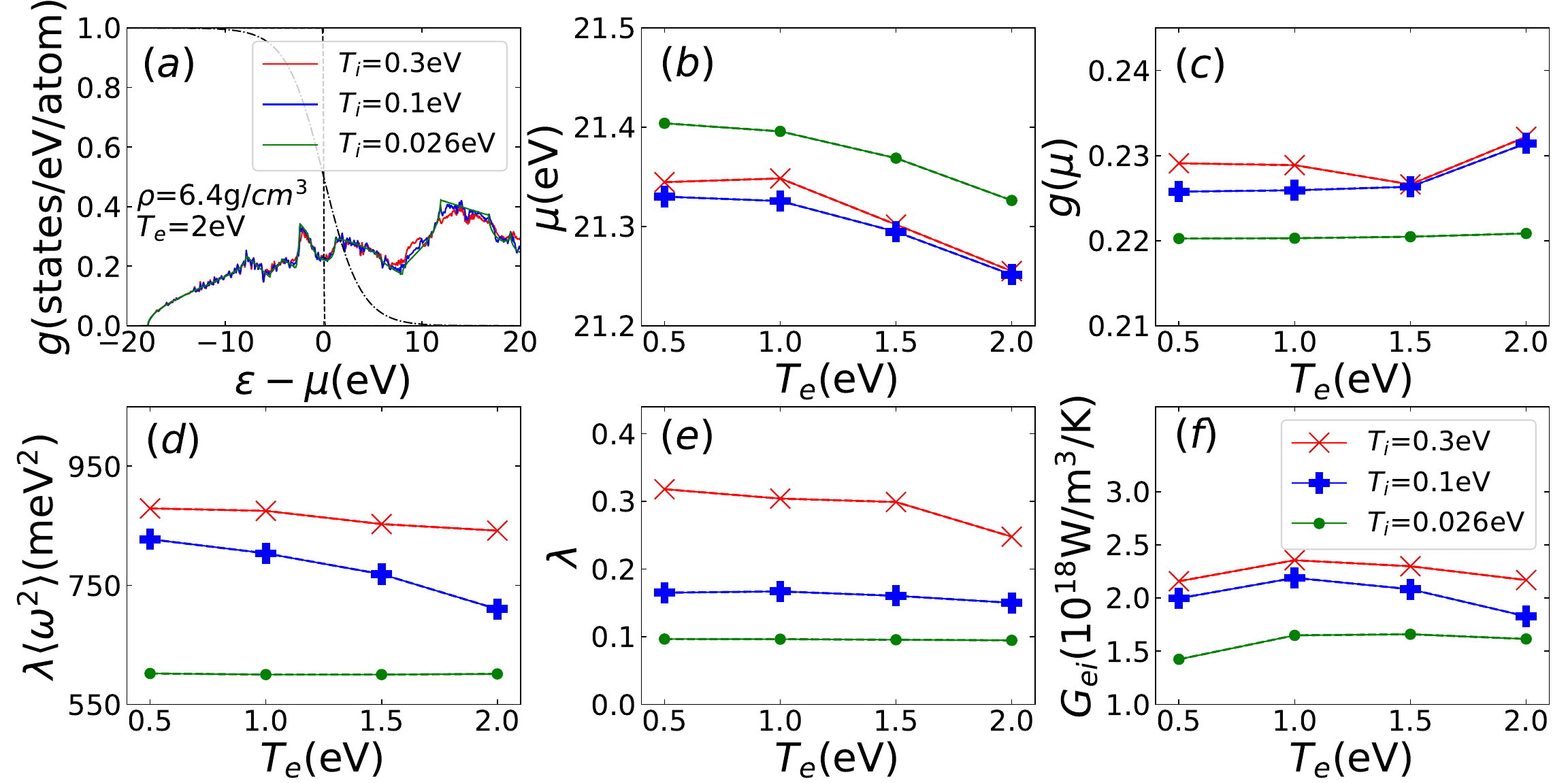}
\caption{The electronic structure and energy transfer rate of warm desne aluminum at  density $\rho=6.4$~g/$cm^{3}$. (a)The electron density of states for three different ion temperatures at high electronic excitation. The two black dashed lines are the Fermi distribution functions at electron temperatures 0.026~eV and 2.0~eV. 
 (b) The chemical potential and (c) the electron density of states at the chemical potential as a function of electron temperature at elevated ion temperatures. (d) The second moment of Eliashberg functions and (e) the electron-ion coupling constants with increasing electron temperature as well as elevated ion temperatures. (f)The temperature-dependent electron-ion coupling factors.}
\label{Figure_10.pdf}
\end{figure}

\begin{figure}[!htbp]
\centering
\includegraphics[width=0.48\textwidth]{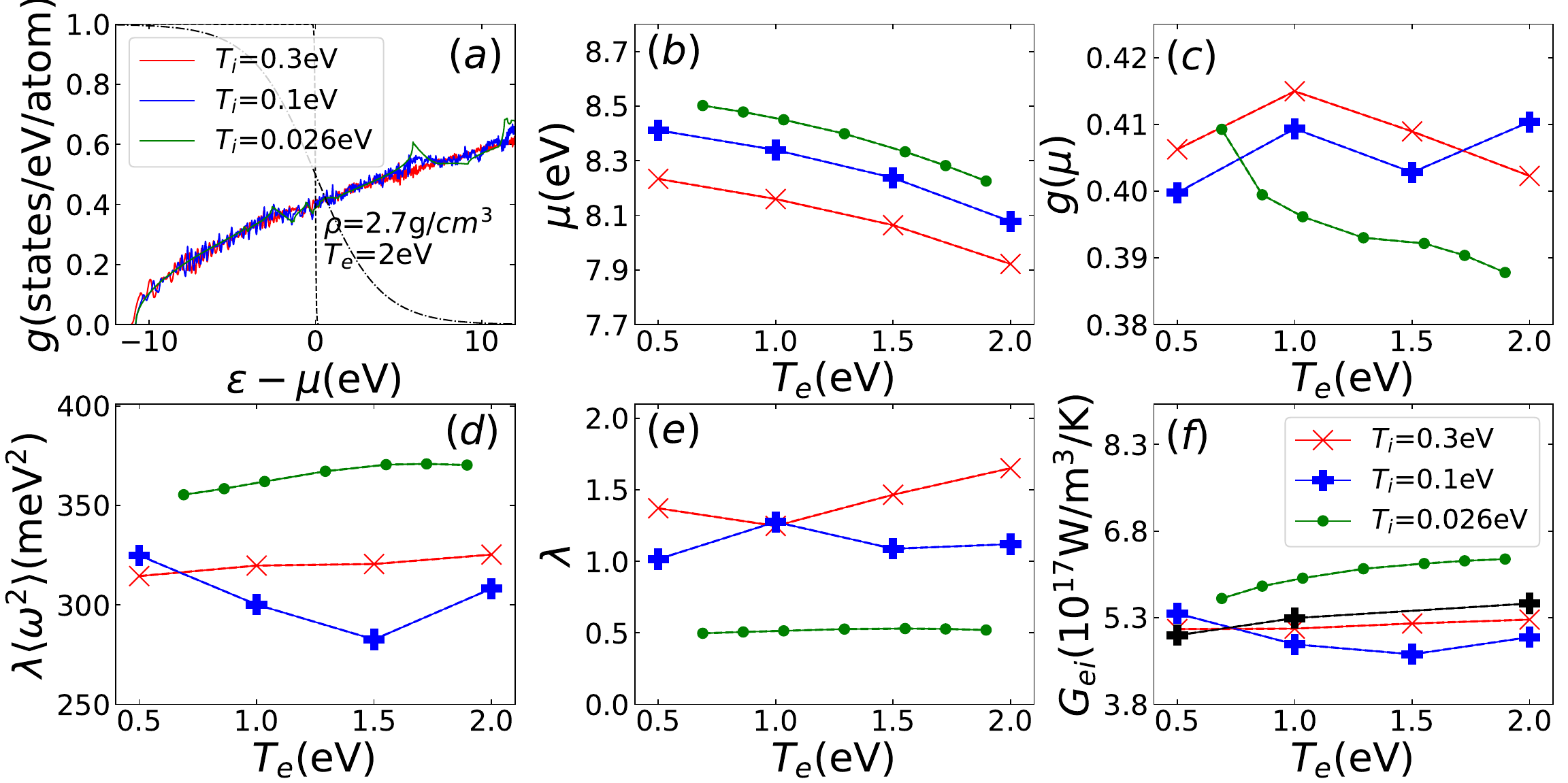}
\caption{The electronic structure and energy transfer rate of warm dense aluminum at  density $\rho=2.7$~g/$cm^{3}$. (a)The electron density of states for three different ion temperatures at high electronic excitation. The two black dashed lines are the Fermi distribution functions at electron temperatures 0.026~eV and 2.0~eV. (b) The chemical potential and (c) the electron density of states at the chemical potential as a function of electron temperature at elevated ion temperatures. (d) The second moment of Eliashberg functions and (e) the electron-ion coupling constants with increasing electron temperature as well as elevated ion temperatures. (f)The temperature-dependent electron-ion coupling factors. All the green curves are taken from our recent work~\cite{zhang2022energy}. The black line in panel f) is adopted from Simoni $et$ $al$.~\cite{simoni2019first}.}
\label{Figure_11.pdf}
\end{figure}

\begin{figure}[H]
\centering
\includegraphics[width=0.48\textwidth]{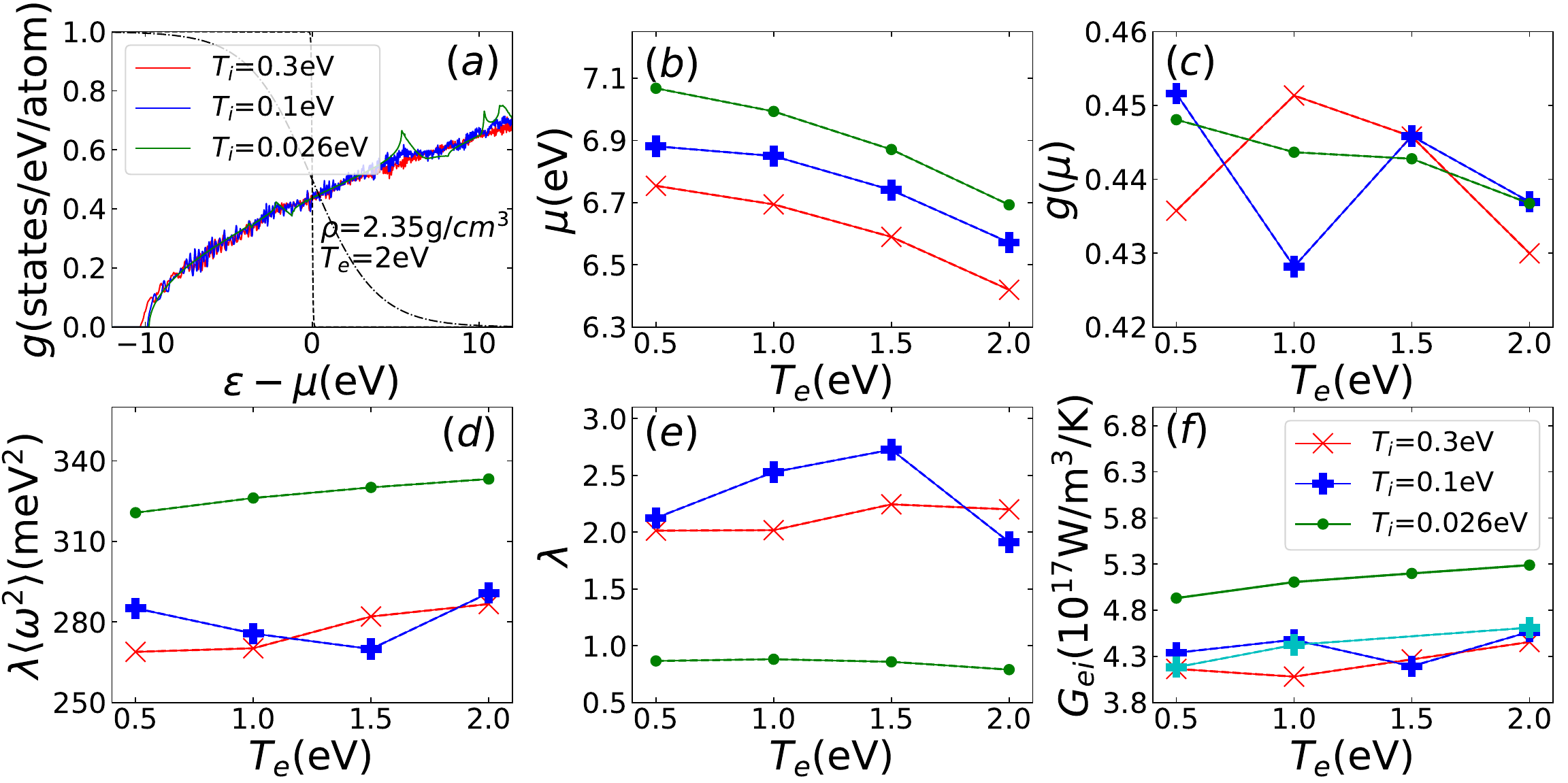}
\caption{The electronic structure and energy transfer rate of warm dense aluminum at  density $\rho=2.35$~g/$cm^{3}$. (a)The electron density of states for three different ion temperatures at high electronic excitation. The two black dashed lines are the Fermi distribution functions at electron temperatures 0.026~eV and 2.0~eV. 
 (b) The chemical potential and (c) the electron density of states at the chemical potential as a function of electron temperature at elevated ion temperatures. (d) The second moment of Eliashberg functions and (e) the electron-ion coupling constants with increasing electron temperature as well as elevated ion temperatures. (f)The temperature-dependent electron-ion coupling factors.
The cyan curve in panel f) is taken from Simoni $et$ $al$.~\cite{simoni2019first}.} 
\label{Figure_12.pdf}
\end{figure}

\bibliography{reference.bib}
\end{document}